\documentclass{an} 
\usepackage{graphicx}
\usepackage{times}
\begin{document}


\title{Astronomical technology -- the past and the future}

\author{I. Appenzeller\fnmsep\thanks{
  \email{iappenze@lsw.uni-heidelberg.de}}\newline}
\titlerunning{Astronomical technology}
\authorrunning{I. Appenzeller}
\institute{Zentrum f\"ur Astronomie der Universit\"at Heidelberg,
Landessternwarte, K\"onigstuhl 12, 69117 Heidelberg, Germany}

\received{2016 Apr 04}
\accepted{2016 Apr 27}
\publonline{2016 Aug 04}


\abstract{The past fifty years have been an epoch of impressive progress
in the field of astronomical technology. Practically all the technical tools,
which we use today, have been developed during that time span. While the
first half of this period has been dominated by advances in the
detector technologies, during the past two decades innovative
telescope concepts have been developed for practically all wavelength 
ranges where astronomical observations are possible.
Further important advances can be expected in the next few decades. 
Based on the experience of the past, some of the main sources of technological 
progress can be identified.}

\maketitle

\section{Introduction}
The topic of this lecture has been chosen for the following reasons: 
Firstly, technology is extremely important
in astronomy. New scientific results quite often result
directly from technical advances. Innovative technologies have also
been among my personal research interests. Finally, this topic is 
rather appropriate for a Karl Schwarzschild lecture,
as Schwarzschild, apart from his fundamental work in astrophysics, 
contributed much to the astronomical technology of his time,
in particular in the fields photographic 
photometry, the understanding of the photographic process,
new observational techniques, telescope optics, and even interferometry. 
A testimony to Schwarzschild's awareness of the important role of 
technology is a lecture on ``Pr\"azisionstechnik und wissenschaftliche
Forschung'' (Precision Technology
and Scientific Research), which he presented in 1914 at the annual meeting
of the German Society for Mechanics and Optics (Schwarzschild 1914). 
In this talk he stated
that ``precision technology is the basis of measuring in time and space, and
on measuring depends, together with the sciences, a major part of our 
whole culture''\footnote{``Die Pr\"azisionstechnik ist ja die Grundlage 
aller Kunst des Messens in Zeit und Raum, und an Ma\ss \ und Messen h\"angt
mit der Wissenschaft zugleich ein gro\ss er Teil unserer ganzen Kultur.''}. 

For this lecture I will not go back to the time of Karl Schwarzschild. 
Instead, I 
will concentrate on 
the past fifty years, as this is the time span where practically 
all the tools have been developed, which we use today. It is also the
period which I could follow personally. 
Because of the scale and diversity of 
astronomical technologies, this report  cannot be complete. The emphasis 
will be on evolutionary landmarks, and there will be a bias
towards optical astronomy.
To provide an idea of the situation fifty years ago, I will start 
with a recollections of my own first encounter 
with astronomy and astronomical technology.    

\section{Becoming an astronomer in the 1960s}
I got involved in astronomy almost by chance while studying physics. 
In the 1960s nuclear physics was most
fashionable. Therefore, after obtaining a degree of ``Vordiplom''
at the University of T\"ubingen in 1961, I moved to G\"ottingen in order 
to do a diploma thesis in the nuclear physics group of Arnold Flammersfeld. 
Unfortunately - or, in hindsight, I would say fortunately -
too many of my fellow students had the same idea. All diploma positions 
at Flammersfeld's institute were filled, and there was a
waiting period of several months. Looking for
alternatives, I found that in astrophysics one could start a thesis
immediately. Thus, instead of working in Flammersfeld's
accelerator lab, I ended up in G\"ottingen's historic astronomical
observatory, where Carl Friedrich Gau\ss \ and Karl Schwarzschild once had
worked.    

The topic of my thesis was a study of the 
the local interstellar magnetic field by measuring  
the interstellar polarization of starlight.  
My adviser was Alfred Behr (1913-2008), who was known
for work on astronomical instrumentation, polarimetry, and 
cosmology. For my observations I was using the historic 34-cm  
Hainberg refractor at the outskirts of G\"ottingen and a polarimeter, 
which Behr had built, and which at the beginning of the 1960s was the most 
accurate instrument of its kind. It could measure stellar linear polarization 
down to about $10^{-4}$. I used Behr's instrument without changes,
but I replaced the mechanical calculators, which at that time were used
for the data eduction, by a computer program. This made it possible to include 
additional terms in the reduction algorithm, which further increased the 
accuracy and reliability of the results. 
The computer, an IBM 650, was owned by one of the local
Max Planck Institutes, but was available to university staff 
when not used by the MPI. It was still
based on vacuum tubes and (compared to modern 
hardware) it was incredibly slow. But it was much faster and 
very much more convenient than mechanical calculators.   
 
Thanks to Behr's excellent instrument
the observations produced interesting data. But due to the often
overcast sky of G\"ottingen, progress was
slow. (According to our national
meteorological service, of all stations with long-term
records, G\"ottingen is one of the three places with the lowest 
percentage of clear sky hours in Germany).  
After watching the clouds above G\"ottingen for about a year, I started 
looking for possibilities to
move Behr's polarimeter temporarily to a better location, and I received 
a positive response from the Haute Provence Observatory 
in Southern France. Behr, my boss, was in the US at that time. So I wrote 
him a letter explaining my plans. He replied that he had an even 
better idea: He had met William Albert Hiltner (1914-1991), 
who had been one of the 
discoverers of the interstellar polarization,  
and who at that time was the director of the Yerkes Observatory of
the University of Chicago. Hiltner had just installed a new telescope
dedicated specifically for polarimetric observations, and he was looking for
somebody helping him to get the new instrument to work. In compensation
for functional work for his institute, Hiltner offered me
a temporary technical position and all the observing time 
I would need to complete my program. Thus, thanks to the 
clouds of the G\"ottingen sky I got the chance to join
- already as a student - one of the major astronomy departments in the 
US. As I learned later, one reason for Hiltner's generous offer
was that Behr had told him about my computer program. Obviously, as soon
as astronomers started using computers, instrument-related software became
a valuable asset.   

\begin{figure}
\includegraphics[width=82mm]{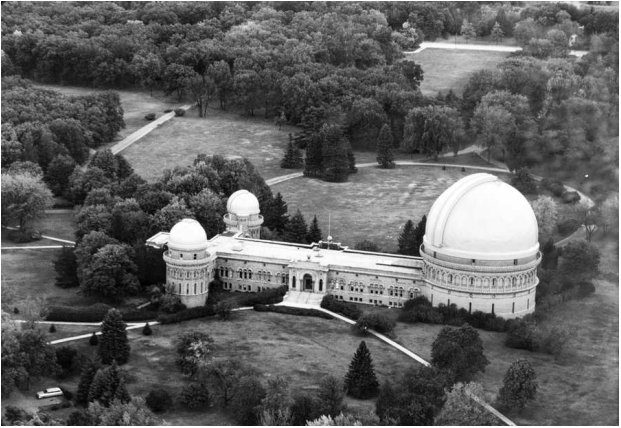}
\caption{Aerial view of Yerkes Observatory. (Image credit: 
University of Chicago Photographic Archive,
Special Collections Research Center, University of Chicago Library).}
\label{yerkes}
\end{figure}
 
In the early 1960s Yerkes Observatory (Fig. \ref{yerkes}), was still
the home the complete astronomy department of the University of Chicago. 
The observatory has been named for the rich Chicago businessman
Charles Yerkes (1837-1905), who had a criminal past and a rather poor
reputation. (More about the interesting CV of Mr. Yerkes can be found in
a history blog by Bruce Ware Allen, http://ahistoryblog.com/2012/08/). 
It was another Chicago citizen, George Ellery Hale
(1868-1938), who converted the poor reputation of Mr. Yerkes into a 
windfall for science. Hale, the first professor of astronomy at
the University of Chicago, convinced Mr. Yerkes 
that he could improve his reputation by financing an 
astronomical observatory for the just founded university. Yerkes agreed 
and told Hale to ``build the largest and best telescope in the world, 
and send the bill to me''. Hale did both,
with enthusiasm. When the observatory opened in 1897, it included 
the famous 40-inch refractor, which is still the world's largest 
lens telescope\footnote{A detailed account of the beginning and the
remarkable history of Yerkes Observatory has been given by 
Donald Osterbrock (1997)}. Hale
later moved to California, where he founded Mount Wilson Observatory
with a 60-inch reflector, to which he later added the 100-inch 
Hooker Telescope. With these instruments completed,  
he initiated the construction of the 200-inch Mount Palomar telescope,
but he died ten years before that telescope could be completed.
At the time of their commissioning all these telescopes were the 
largest functioning instruments in the world. 
Hale also founded or initiated the
Astrophysical Journal, the American Astronomical Society, the US 
National Research Council, and many other important organizations. 
He was arguably the
personality most responsible for the pre-eminence of the US 
in astronomy during the following decades.

During much of the 20th century Yerkes Observatory remained an
important center of astronomical research. While Otto Struve 
was its director or department chair between 1932 and 1950, 
he managed to attract outstanding scientist from all over the world
to Yerkes. Many of the staff members of that time later held 
leading positions in the US and European astronomy, and some played 
decisive roles in the development of modern astronomical techniques. 
By the time when I joined the institute
in summer 1964, Yerkes Observatory had already passed its 
apogee, but its staff still included first-class scientists, 
such as Subrahmanyan Chandrasekhar, W. W. Morgan and W. A. Hiltner. And
Yerkes was still connected with the McDonald Observatory in West Texas, 
which belonged to the University of Texas, but which was founded
by Otto Struve, and initially operated by Yerkes staff, 
until the University of Texas could establish an astronomy department 
of its own.

Hiltner's new polarimeter followed the same principle as
Behr's instrument in G\"ottingen. In both cases a 
Wollaston prism was used to split the incoming light 
into two perpendicularly polarized beams, which were directed
to two photomultiplier tubes (PMTs).  
The amount and position angle of the linear polarization 
of the incident light was derived by rotating the
Wollaston prism together with the detectors (in steps) around the 
optical axis. For linearly polarized light this results in a
periodic variation of the difference (and ratio) of the two output beams.
Fourier-analyzing this difference signal gave the amount and 
position angle of the polarization. In G\"ottingen 
the polarimeter was rotated behind the telescope. In this case the
measured polarization is a superposition of the stellar polarization
and any polarization produced by the telescope optics. Unfortunately,
telescopes tend to produce instrumental polarization, which often is variable. 
Therefore, at Yerkes the
the polarimeter was rigidly attached to the telescope and
the whole telescope tube, including the optics, was rotated together 
with the polarimeter. In this case the instrumental polarization 
produces  a constant intensity difference between the two beams only, which 
does not affect on the measured polarization. 
Together with a similar telescope, which was installed at the  
Siding Spring Observatory in Australia, the Yerkes instrument was used to
set up a system of polarimetric standard stars with well established 
polarization parameters that were free of instrumental polarization. 

\begin{figure}
\includegraphics[width=82mm]{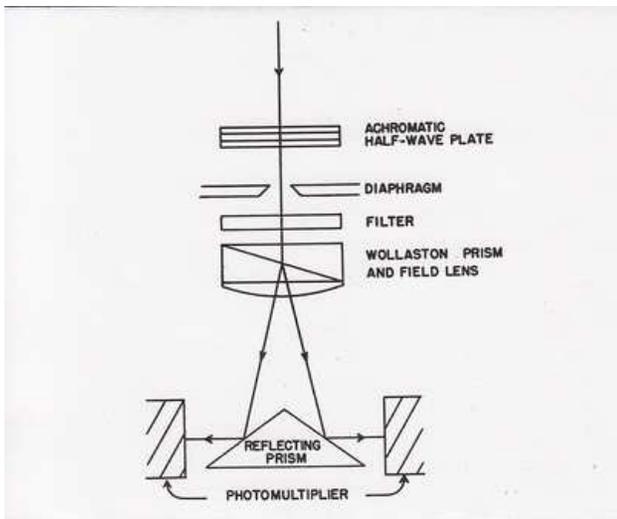}
\caption{Optical layout of the half-wave-plate polarimeter built
for Yerkes Observatory. Details see Appenzeller (1967).}
\label{halfwave}
\end{figure}

A disadvantage of Hiltner's new instrument was that
the rotation of the telescope and the need to re-center the
target at each rotation step resulted in a relatively low ratio between 
the on-target integration time and the total time required for
each observation. In order to reduce the overhead, I suggested 
to Hiltner to keep the telescope and 
polarimeter fixed and to rotate the plane of vibration of the incident 
polarized light. As described in the textbooks of optics, this
can be done by means of a half-wave plate.
For broad-band applications an achromatic half-wave plate 
is required, which can be constructed by combining
three normal wave plates with different orientations of their optic axes, 
as had been shown by the Indian optician Shivaramakrishnan 
Pancharatnam (1955).   
Hiltner liked the idea and asked me whether I could build a 
half-wave-plate polarimeter
for his institute. That request was the beginning of my
career in building astronomical instruments. 

Being still a student
at G\"ottingen, I could not start the new instrument immediately,
as I  first had to go back to Germany  for about a year to finish my studies. 
But in the spring of 1966 I was back in Chicago and started constructing
the half-wave-plate polarimeter (Fig. \ref{halfwave}). 

During the year in G\"ottingen I visited 
the 1965 fall meeting of the Astronomische Gesellschaft in Eisenach,
Th\"uringen,
where I gave my first (short) talk at an AG meeting. That meeting took
place almost exactly 50 years ago. Thus, for me the talk today, is, in 
a sense, a semicentennial.
With this remark I conclude the personal part of this talk, passing on
now to a more general overview of the last fifty years.

\section{The 1960s: Big discoveries, small technical advances}

Scientifically the period 1960-1969 was an exceptionally exciting time. 
During that decade
probably more new types of astronomical objects were discovered than in any 
ten-year period before or after. 
Up until 1960 astronomy was essentially restricted to studies of the
solar system, the stars, galaxies (including our own), and 
the interstellar matter. The existence
of dark matter had already been shown, but was largely ignored. By 1970 
we knew more than twice as many types of astronomical sources. Among the
newly discovered species were the stellar X-ray sources (found in 1962),
the quasars (1963), the pulsars (1967), the cosmic microwave background (1965),
and the gamma-ray bursters (first observed in 1967). Moreover, during that 
decade the first FUV radiation, X-rays, and $\gamma$-rays from the 
Milky Way and from other galaxies  were recorded (1967, 1968), and
man landed on the moon (1969). These discoveries greatly influenced astronomy, 
but most were based on conventional technologies. The landmark 
spectroscopic observations still used photographic plates.
The X-ray observations were made with old-fashioned Geiger counters, 
and the CMB was discovered with an antenna that 
had been built years earlier for communication purposes.

There was some progress in 
the detector field. For the first time image intensifier 
tubes were used with photographic plates, and
Frank Low (1933 - 2009), who worked at Texas Instruments
at that time, invented the germanium bolometer (Low 1961), which
was much more sensitive and covered a much broader wavelength
range than earlier IR detectors. Starting 1965 he used his new detector
to carry out the first infrared observations from above the
atmospheric H$_{2}$O (first on a military airplane, operated by
the US Navy, then with a Lear Jet of NASA), 
setting the stage for the later Kuiper and SOFIA stratospheric 
observatories (see, e.g., Aumann, Gillespie \& Low 1969). First 
steps towards space astronomy were the launches of the solar
observatories OSO 1 to 6, and the UV satellite OAO 2, which
included UV-sensitive TV cameras and PMT-based UV-band photometers. 
Among the results of OSO 3 was the first detection of $\gamma$-photons
from space.

An important step for optical photometry was the introduction
of photon counting with photomultipliers.
Counting individual visual-light photons 
was, in principle, possible since photomultipliers had been invented. 
The internal amplification of the PMTs (of the order $10^{7}$) produces 
charge pulses that could be individually detected, even with the noisy
vacuum-tube electronics of the 1960s. Photon counting was well known 
to give the maximal attainable photometric accuracy, limited only by
the statistical variation of the number of photons received per unit of time. 
Several astronomers, including Hiltner, 
had been experimenting with the photon-counting technique in the
1950s, but had given up because of non-linearities. 
While all counting methods become nonlinear due to dead-time
effects at some rate, in the case of photomultipliers
nonlinearities connected with the peculiar pulse shape of PMTs 
set in already at relatively low light fluxes. Therefore, in the early
1960s photometry (and polarimetry) was done by measuring the DC
current produced by PMTs. However, this results in an additional statistical
error due to the varying charge multiplication factors of the individual
photo electrons. Therefore, in the DC mode the {\it detective} quantum 
efficiency of PMTs is much lower than the {\it responsive} QE of the cathode. 
Among the work which I did at Yerkes was developing an electronic circuit 
which avoided the nonlinearities in photon counting, and
the half-wave-plate polarimeter, which I built there, 
used such an improved photon counting system. As we had the
DC amplifiers still at the telescope, we could carry out a direct 
comparison of the two methods under identical conditions.  As expected
from the theory, photon counting reduced the statistical errors by
factors of about 2 to 3, and made it possible to observe faint objects,
where the DC method produced noise only.

Another innovation of the new polarimeter was a programmable 
controller (a small custom-built process computer), which
operated the half-wave plate, started and stopped
the integrations, and initiated the data readout and storage.
The automatic control, the facts that a target re-centering 
was no longer required, and that the sky background observations could be
carried out in a more flexible way, reduced the 
the observing time for a given object by factors
between 5 and 10. A  program which before required about  
one week could now be done in one night. 
 
As seen from today, the Yerkes polarimeter was an example of the 
minor technological improvement of existing technologies in the 1960s,
which nevertheless resulted in a significantly better performance 
and in a much more efficient use of the valuable observing time.  

\section{The 1970s: Photocathodes replace photography, Einstein, COS-B, 
and the MMT}
While the technological advances of the 1960s were moderate,
the pace of technological progress increased dramatically after 1970.
Most important was 
the replacement of the photographic plates by photocathode-based
2D detectors. Among the 
first and most successful examples of these new detectors was the
image-dissector scanner developed at the
Lick Observatory (Robinson \& Wampler 1972). But similar 
devices were built simultaneously  
at many other places in the US and in Europe. Their use in space begun  
with the International Ultraviolet Explorer (IUE) satellite, 
which was launched in 1978. Due to their 
higher quantum efficiency, photocathode detectors were faster and reached
fainter sources.  But more important was their linearity, 
which (in contrast to photography) made it possible
to subtract sky backgrounds. This allowed us for the first time 
to obtain  quantitative images and spectra of targets whose 
surface brightness was well below that of the night sky. 

\begin{figure}
\includegraphics[width=82mm]{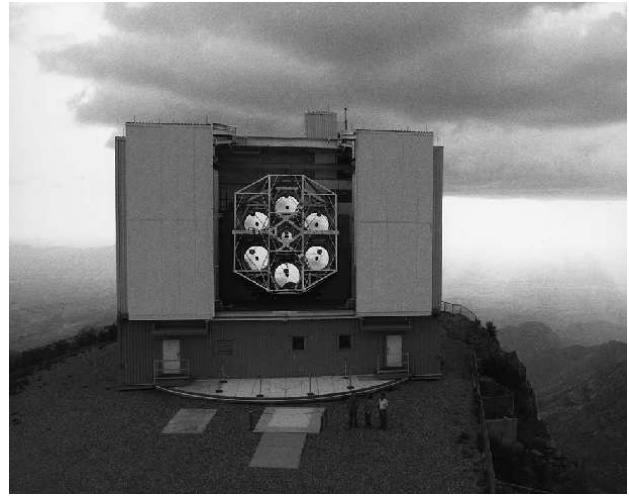}
\caption{The original Multiple Mirror Telescope on Mount Hopkins, Arizona,
consisting of six 1.8-m mirrors in a common mount and with a common focus.
Photo: Smithsonian Institution Archives.}
\label{mmt}
\end{figure}

A highly important event of the 1970s  was the completion of the   
Multiple Mirror Telescope (MMT, Fig. \ref{mmt}) on Mount Hopkins in 
Arizona in 1979 (Beckers et al. 1981). 
Its concept had been developed by Aden Meinel (1922 - 2011) with support by
Frank Low, and Fred Whipple (Meinel et al. 1972). While Hale clearly
was the leading personality for the development of astronomical telescopes
in the first half of the 20th century, Aden Meinel arguably played this role 
after 1950. He was born in Pasadena, and, before studying optics and 
astronomy at Pasadena and Berkeley, he worked 
as an apprentice in the optical lab of the Mount Wilson Observatory. During 
the Second World War he was involved in developing rockets for the US navy. 
Between 1949
and 1956 he worked at Yerkes Observatory (after 1953 as 
its deputy director), before being appointed the
first director of the National Optical Astronomy Observatory (NOAO), which
became Kitt Peak National Observatory. Meinel later headed the Steward
Observatory of the University of Arizona, and then founded the
Optical Science Center of the UoA, which developed into today's 
prestigious College of Optical Sciences. After retiring from the UoA he worked
at the Jet Propulsion Laboratory in his home town Pasadena on  
the ``Large Deployable Reflector'' project, which 
developed into the James Webb Space Telescope. 

Meinel realized that with the Palomar telescope the large rigid 
mirrors and the equatorial mounts had reached a dead end. 
Already at Yerkes Observatory Meinel developed concepts for 
innovative new telescopes,
including a 400-inch Arecibo-type optical instrument
(see Meinel 1978). At NOAO he 
initiated the ``Next Generation Telescope'' study program,
which discussed the possible (and impossible) telescope concepts
promoted at that time. At the same time he established at NOAO
a working group to plan a large optical space telescope. 
Together with the work of 
a similar study group initiated by Lyman Spitzer (1914 - 1997) 
at Princeton University  Meinel's group was later merged 
into a NASA project, which eventually resulted in the Hubble Space 
Telescope. Although after 1946 he worked at civilian 
institutions, Meinel never completely cut his relations with 
the US military. In this context he 
learned about six high-quality 1.8-meter mirror blanks, made by Corning, 
which had been ordered, but never been called for, by the US Air Force.
Meinel acquired these blanks for a symbolic price and constructed a
telescope with the six mirrors in a common mount and with a common focus
(Fig. \ref{mmt}).
After some initial difficulties, the Multi-Mirror Telescope
(with the light collecting power of a single 4.5-m aperture) worked quite
well. It was the first segmented reflecting telescope and the first telescope
depending entirely on active optics, and it demonstrated that 
instruments with large effective apertures could 
be realized in this way. Thus, the MMT
formed the basis of the new-generation large optical/IR
telescopes of the 1990s and beyond.  

\begin{figure}
\includegraphics[width=82mm]{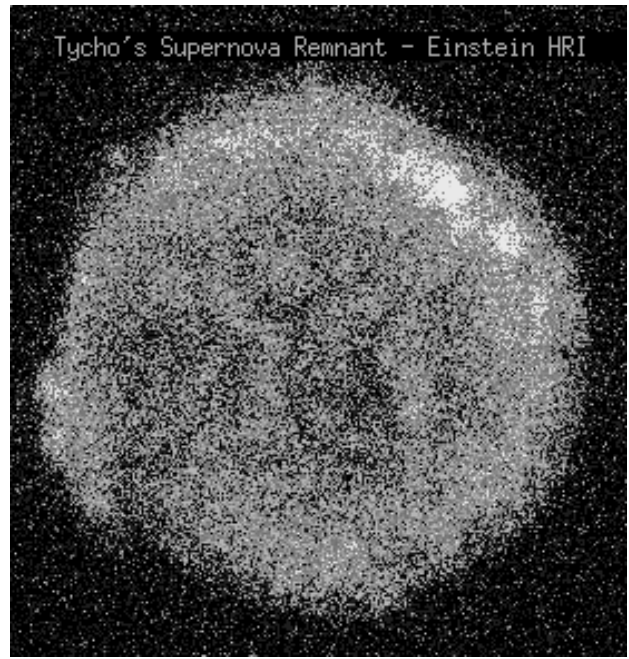}
\caption{X-ray image of the Tycho supernova remnant obtained with the
HEAO-2 (Einstein) observatory. With an angular resolution 
reaching about 2 arcsec, HEAO-2 was the first mission capable 
of high-resolution X-ray imaging. (Credit: NASA).}
\label{einstein}
\end{figure}

In the field of X-ray astronomy a major milestone was 
the first (still modest) X-ray sky survey by Riccardo Giacconi and his group
using the Explorer 47 (Uhuru) satellite, launched in 1970. 
Technologically more important was the
HEAO-2 (Einstein) X-ray observatory, launched in 1978. Its 
payload included the first
large grazing-incidence telescope, as well as imaging
X-ray detectors, which produced first high-resolution
X-ray images of astronomical objects. An example is shown in 
Fig. \ref{einstein}. 
A big step forward in our knowledge of 
the gamma-ray sky was the launch of the COS-B survey satellite in 1975.

In addition to the MMT, the 1970s witnessed the construction of 
seven new optical 
telescopes with apertures larger than 3 meters. But, apart from
the altazimuth mounting of the Russian 6-m telescope, 
these instruments were still essentially based on concepts and
technologies developed before 1960. The same was true for the (scientifically
highly successful) OAO 3 (Copernicus) FUV space mission. 
    
More innovative was the new 100-m telescope of the Max Planck Institute
for Radio Astronomy at Effelsberg, which was completed in 1971. Its
overall design followed the conventional ``dish'' concept, pioneered by
Grote Reber in 1937.
But, as an important new feature, the mechanical structure
of the Effelsberg dish was designed to always retain
a paraboloid surface in spite of the unavoidable
deformation during altitude changes. This ``homologous'' design 
concept made it eventually  
possible to use the 100-m dish at wavelength as small as a few
millimeters.

\section{Between 1980 and 1992: CCDs, the VLA, 
space observatories, and VHE gamma photons}

Although they had been developed with
much effort only in the decade before, 2-D photocathode devices
became obsolete when CCDs and other solid-state array detectors became 
available during the 1980s.
The CCD detector was invented at the Bell Labs in 1969
as an imaging device for ``picture phones''. The first ones 
were rather noisy, and by 1975
their usefulness for astronomy was still disputed. But,
in the following years low-noise CCDs were developed, and already
in 1976 a first astronomical application was reported at
an AAS meeting (Loh 1976). At the 1981 SPIE meeting on ``Solid State
Imagers for Astronomy'' (Geary \& Latham 1981) various US and 
European groups reported 
their positive experience with the new detectors, and during the
following years CCDs and NIR array detectors were
installed at all major observatories, while the photocathode detectors 
disappeared, except for some UV and soft X-ray applications.

What made the semiconductor array detectors so attractive, was their 
high quantum efficiency (reaching close to 100 per cent), their
mechanical robustness, their lower sensitivity to overexposure, and  
the fact that (in contrast to photocathode devices) no high 
voltage was required.

\begin{figure}
\includegraphics[width=82mm]{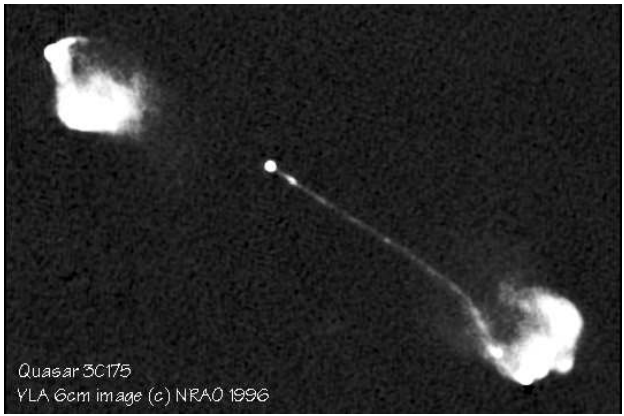}
\caption{Radio image of the Quasar 3C175 with a resolution of about 350 mas,
obtained with the Very Large Array. The field is about 1.2 $\times$ 0.6
arcmin. (Credit: NRAO).}
\label{vlaim}
\end{figure}

Another landmark in the development of astronomical imaging was the
completion of the VLA in 1980. A directional antenna system had already 
been used by
Jansky in his discovery of radio radiation from space, and 
first radio interferometers had been constructed during
WW2 for military purposes. In the years after the war large astronomical
interferometers had been installed in England, Australia, and the Netherlands. 
But it was the size and the non-redundant design of the VLA 
that made it possible for the first time
to produce routinely complete and well resolved radio images of complex
astronomical sources, as illustrated in Fig. \ref{vlaim}. These 
observations demonstrated 
convincingly the imaging capabilities of aperture synthesis, which had
a major influence on the telescope design at radio, IR, and visual 
wavelengths during the following decades.

Among the advances in space astronomy was the 1983 launch
of the IRAS satellite. This instrument, which had been proposed by 
Frank Low and realized by a US-European cooperation, 
opened up the FIR sky. 

Even more important was the Hubble Space Telescope, launched in 
1990, which was scientifically particularly successful and which 
became the most popular NASA mission to date. Among other results, the HST 
dramatically extended the redshift range at which galaxies and QSOs 
could be observed. 

During the same year the launch of the ROSAT satellite resulted in a
first inventory of the cosmic X-ray sources. 
 
Less widely known, but not less important was the Compton
Gamma Ray Observatory (launched in 1991) which produced (among other results)
the first sky map
for photon energies above 100 MeV and provided critical new information
on the gamma-burst sources. 

A smaller, but technically highly innovative space mission was
the Hipparcos satellite (launched in 1989) which revolutionized
the field of astrometry. 
 
At the early space observatories (such as the highly successful
IUE satellite) the observations were carried out in the same way 
as the work in ground-based observatories of that time, 
with a single scheduled observer operating the telescope and the
instrumentation. The only difference was that the individual steps 
were carried out by remote control. The Hubble Space Telescope 
was the first observatory
which was fully based on computer-controlled automated instruments 
and queue scheduling. The advantages and economy of this operating mode soon 
became evident, and within a few years these techniques became routine 
at practically all major ground-based facilities as well.

Other technical breakthroughs of the 1980s were the first astronomical
use of adaptive optics (originally developed for military applications) 
and the first detection (in 1989) of TeV photons (from the Crab nebula) 
with the air-shower Cherenkov telescope of the
Whipple Observatory on Mount Hopkins in Arizona. With this addition, 
the wavelength range
of electromagnetic radiation available for astronomers reached the present-day 
interval of about $10^{-20}$ m to about 20 m (a frequency range of 
about 70 octaves!).
 
\section{Starting 1993: A new generation of large telescopes}
While the period 1960 - 1990 was dominated by advances in  
detector technologies, the last two decades were marked    
by the completion of innovative new telescopes.

The construction of these new large facilities extended to all
astronomically useful wavelength ranges and involved institutions
in many different countries from all inhabited continents.  

\begin{figure}
\includegraphics[width=82mm]{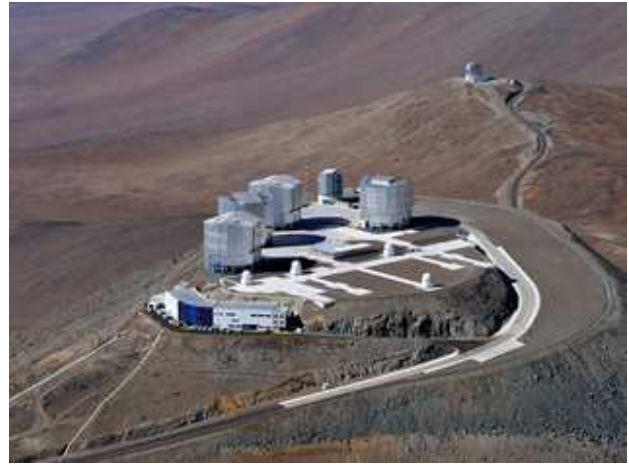}
\caption{The Very Large Telescope (VLT) of the European Southern Observatory
at Paranal, Chile. The light from the four large (8.2-m) telescopes
and the four smaller (1.8-m) telescopes can be combined coherently
to reach an angular resolution at visual/NIR wavelengths 
of about one milliarcsecond. (Credit: ESO).}
\label{vlt}
\end{figure}

At optical/IR wavelengths the important first example was the 
Keck-1 telescope with its 10-m segmented mirror, completed in 1993
(Nelson \& Gillingham 1994). During
the following years 15 more telescopes with apertures above 6 meters 
and actively controlled and/or segmented mirrors were installed
in different parts of the world. 
Apart from a superior light collecting power of these large facilities, their 
actively controlled optics resulted in a significantly 
better image quality, even for seeing-limited observations. Some
of these new instruments, such as the ESO-VLT (Fig. \ref{vlt}), LBT, and Keck
1+2) were designed to
include interferometric and aperture synthesis capabilities.
Although less big, the astrometric satellite GAIA (launched in 2013) 
has also to be mentioned, as its will a have a big impact on 
several subfields of astronomy.

For shorter wavelengths a new generation of advanced X-ray telescopes,
such as ASCA (1993), Chandra, and XMM-Newton (both launched in 1999) 
became operational.
Using large grazing-incidence optics, X-ray CCDs with a spectral
resolution of about R=50, and grating spectrometers for higher
spectral resolutions, the data produced by these X-ray 
observatories were for the first time fully comparable 
to high-quality optical data.

\begin{figure}
\includegraphics[width=82mm]{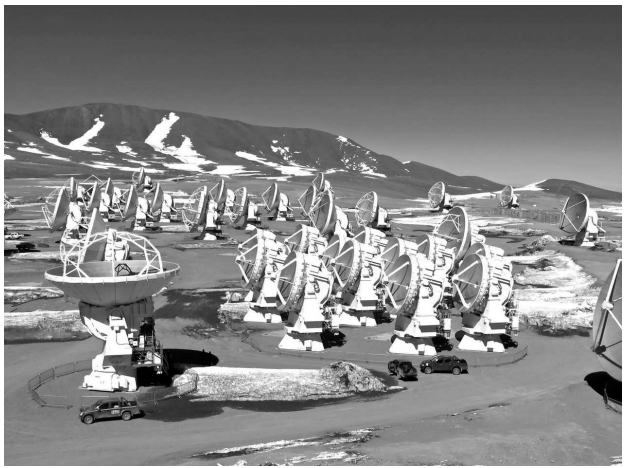}
\caption{The Atacama Large Millimeter/submillimeter Array (ALMA) on 
the Chajnantor plateau, Chile. (Credit: ESO).}
\label{alma}
\end{figure}

FIR observatories with actively or passively cooled space telescopes, 
such as ISO (1995), Spitzer (2003), and Herschel (2009) resulted in
a significant improvement of the sensitivity limit at these wavelengths.
A similar role was played for the EUV by the FUSE observatory (1999),
for the gamma-ray astronomy by the INTEGRAL (2002) and Fermi (2008)
space missions, 
and for the VHE gamma range by the new large Cherenkov air shower
telescope systems HEGRA, H.E.S.S., and VERITAS. Finally, the
completion of the ALMA array in Chile in 2013 (Fig. \ref{alma}) resulted in 
a major improvement of the the sensitivity,
resolution, and imaging capabilities at millimeter 
and sub-mm wavelengths. 

In addition to the multi-purpose instruments listed above, some
important ``single-target'' telescopes were realized. Those
included the WMAP and PLANCK space missions (for studying the CMB) 
and new large solar telescopes,
such as the NST, installed at the Big Bear Lake Observatory, California, 
and the GREGOR telescope, constructed on the Spanish island of Tenerife.  

\section{The future} 
Many of the facilities listed in the last section are still in operation.
Thus, this overview  has reached the present-day status, and we may
now ask what kind of technological progress the future may have in store.
Such an outlook is necessarily speculative. However, there are
at least three fields where predictions appear reasonably safe:

\subsection{Projects in progress}
At present several large  
projects are already under construction or in an advanced planning stage. 
These will almost certainly dominate the technological
efforts for at least another decade. They again cover the whole
wavelength range between the radio and the VHE-gamma energies.
Although their realization is not yet in all cases assured, at least most of 
these projects will probably become reality.

Closest to completion are the Astro-H X-ray facility
 and the James Web Space
Telescope for very deep NIR observations. Both are expected 
to be launched before the end of the present
decade. During the following decade three big ground-based optical telescopes
(the Thirty Meter Telescope, the 39-m ESO E-ELT, and the 21.4-m equivalent
Giant Magellanic Telescope),  the radio-frequency Square Kilometre Array, 
the Cherenkov Telescope Array for the VHE gamma astronomy, and (possibly)
ESA's X-ray telescope ATHENA may receive their first photons. 
Although all the new instruments promise great scientific progress, 
from the standpoint 
of technology they are in most cases less exciting, as they are mainly
extrapolations of existing facilities to much larger sizes. Exceptions are 
ATHENA with its innovative silicon pore optics and Astro-H with
new superconducting detectors.

A smaller, but scientifically not less important project, is the 
Large Synoptic Survey Telescope (LSST). Apart from tracing all types 
of variable sources,
the very deep new map of the sky, which the LSST will produce over
the years by coadding frames, will be a most valuable data base for
many future scientific studies. Among the challenges connected with the
LSST will be the software tools needed to extract information
from such large data bases efficiently. 
  
The new instruments mentioned above will bind much of the financial and
personal resources available in the near future. Thus, 
it may be prudent to wait for the experience with 
these projects, before
planning an even more advanced generation of telescopes.
On the other hand, one should not stop thinking about the
distant future, although it may take some time before 
dreams such as observatories on the moon may become a reality.

\subsection{New photon detectors}
On shorter timescales we may well experience important progress in
the fields of focal plane instrumentation and, in particular,  
energy-resolving photon detectors.
For any individual photon the frequency $\nu $ and wavelength $\lambda $ are
well determined by the relation $E = h \times \nu =h \lambda ^{-1} c$ 
where E is the photon
energy. However, at present only at gamma and X-ray wavelengths the
photon energies (or wavelengths) are determined directly by 
the detectors, while
at longer wavelengths more or less complex spectrometers are used 
to sort the photons before they are detected. 
Obviously this not very satisfactory situation is due to the use of  
semiconductor detectors and the fact
that the most common semiconductors (such as silicon) have intrinsic
band gaps corresponding to about the energy of visible photons. 
Hence, visible photons can produce single photo electrons only, 
while the more energetic photons can produce  
multiple conduction electrons, where the number (or total charge) can
be used to derive the energies of the individual detected photons. 

It has been known since many years,
that in superconducting metals below a critical temperature band gaps
are formed which are much narrower (typically 0.1 - 1 meV). Thus, 
superconducting detectors can measure photon energies
with an energy resolution for visible light, which is comparable to that of 
Si-based CCDs for X-rays. In a similar way superconducting 
(micro-) bolometers can be used as photon-energy recording detectors.
Bolometers (where photon energies are determined
indirectly by measuring the resulting heat input), can
be used at all wavelengths, where photon absorption is possible,
including the UV, visual, IR and sub-millimeter ranges.
Single-pixel superconducting photon detectors and small 
arrays based on these principles have been constructed and tested since 
more than 15 years (e.g. Jakobsen 1999). But, most of these devices are 
limited to pixel formats which are not competitive with modern CCDs. A 
possible breakthrough could be the
Microwave Kinetic Induction Detector (MKID) read-out principle.
MKIDs are already used at mm- and sub-mm observatories and first
scientific results obtained in the visual have been published. The
presently existing visual-wavelength MKID arrays still have a few 
thousand pixels and a visual-light spectral resolution of 
R=$\lambda/\delta \lambda \approx 10$ only, but megapixel arrays
and spectral resolutions of the order $10^{2}$ appear to be 
technically feasible (Mazin et al. 2013, 2014). If this can be achieved,
MKID-based photon detectors may soon replace the CCDs for low-resolution
imaging spectroscopy and many other applications. 
The low operating temperature (of the order 100 mK) of the 
superconducting devices results in
a more complex operation, but advances in the low-temperature technologies
make such systems increasingly feasible and affordable.

Energy-resolved photon counting and imaging using        
microbolometers (also known as microcalorimeters) have
also been introduced in X-ray astronomy, where they reach  
a significantly better spectral resolution than is possible with silicon CCDs.
However, as X-ray observatories have to be operated in space,
the low detector temperatures are even more challenging.    

\begin{figure}
\includegraphics[width=82mm]{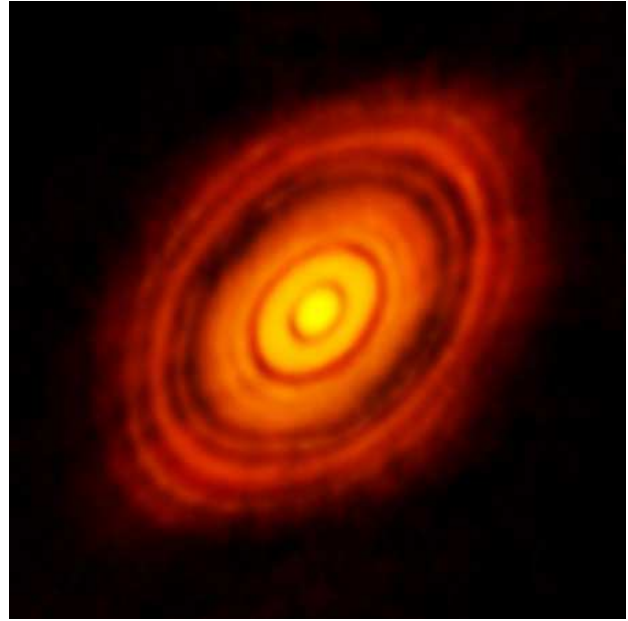}
\caption{Millimeter-wave image with a resolution of about 50 mas of 
the circumstellar disk around the very young (T Tau type) star HL Tauri 
obtained with the Atacama Large Millimeter/submillimeter Array.
(Image credit: ESO.)}
\label{hl}
\end{figure}

\subsection{Aperture synthesis}
Another technique where important further progress can be safely predicted,
is the field of interferometry and aperture synthesis. The big success
with radio interferometers has already been mentioned. More
recently ALMA has demonstrated the great potential 
of such techniques at mm and sub-mm wavelengths. Circumstellar disks
around young stars, which 30 years ago could be inferred only indirectly,
can now be imaged and studied in significant detail (Fig. \ref{hl}). 
At NIR wavelengths observations with the VLT and the Keck 
interferometers already resulted in very valuable scientific data, 
and the LBT promises new possibilities in the interferometric imaging
of complex objects. As an example I note the imaging of the 
symbiotic binary system SS Lep, obtained in the H-band by Blind et al. (2011)
with the VLTI/PIONIER instrument. In view of the ongoing
technical developments, we can expect that this type of work will soon be
expanded to a wider wavelength range and to sources with more complex
geometries.    

\section{The sources of technological progress}
Looking back on the history of the past five decades, 
it appears to me that the main sources of major advances are

\begin{itemize}

\item[(1)] a steady improvements of existing technologies,

\item[(2)] the adaption of new technologies developed in other fields,

\item[(3)] outstanding personalities with a bold vision and the political 
           skills to realize new big projects, and

\item[(4)] competent and motivated teams of scientists, engineers,
           and technicians, who are dedicated to their work, and
           who enjoy technological challenges. 

\end{itemize}

Steady improvements of existing
technologies by new ideas and the adoption of new techniques
developed elsewhere does not sound very exciting. But much of 
the past progress was actually based on such gradual improvements.
Examples are Schwarzschild's diligent work on the photographic
methods, but also the improvements of the existing techniques
in the 1960s described above. Among the important technologies 
adopted from other fields are the telescope, photography, photomultipliers, 
computers, CCDs, and adaptive optics.    

Large projects, such as innovative new 
telescopes, new wavelength ranges, and completely new methods,
can often be traced to the vision, the tenacity, and 
the political skill of outstanding personalities. 
Examples which I mentioned already  
are George Ellery Hale, Otto Struve, Aden Meinel, Frank Low, and
Lyman Spitzer. Other examples are Riccardo Giacconi, who initiated and
dominated the development of X-ray astronomy, David Heeschen, who 
was the main driver for the VLA project, and Yasuo Tanaka, who pioneered
the use of energy-resolving CCDs in X-ray astronomy.  European colleagues,
who played such roles, are, e.g., Martin Ryle, 
the founding fathers of ESO (including Jan Oort, Otto Heckman,
and Andr\'e Danjon), Lodewijk Woltjer, who initiated ESO's Very Large
Telescope. Examples in Germany are Karl-Otto Kiepenheuer, 
Otto Hachenberg, Joachim Tr\"umper, and Hans Els\"asser. 

\begin{figure}
\includegraphics[width=82mm]{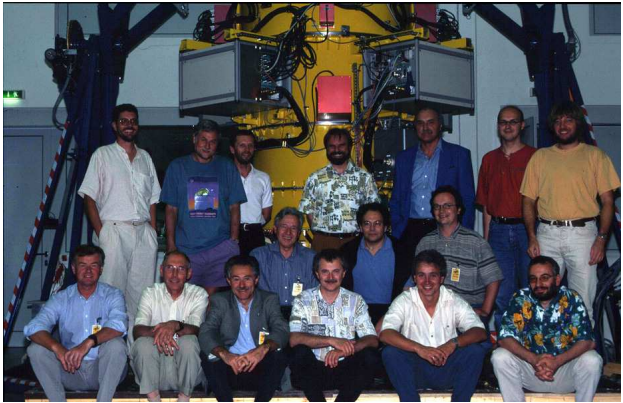}
\caption{Part of the team which developed the FORS instruments for
ESO's Very Large Telescope. The photo was taken during the acceptance
tests of FORS1 in Europe. FORS1, which was the first science instrument 
at the VLT, is visible in the background.}
\label{fors}
\end{figure}

An interesting document in this context is a staff photo taken
at Yerkes Observatory in 1950. This historic photo includes 
a sizable fraction of the (US and European) ``visionaries'' which I 
mentioned above. (The Photo can be found in the photographic archive 
of the University of Chicago, which is accessible in the 
Internet through the UoC web page).  

Finally, no major technological project can be successful without the
cooperative work of motivated teams of scientists, engineers,
and technicians, who are dedicated to their work and who enjoy
technological challenges. Fig. \ref{fors} shows part of such a team,
with which I had the pleasure of realizing a successful
astronomical instrument. The group includes scientists as well as 
technical experts of various fields. That highly 
motivated teams of scientists {\it and} technicians are 
indispensable for such work, was again stressed already by 
Karl Schwarzschild in his lecture
to the German Society for Mechanics and Optics. There he
characterized the ``right'' scientists {\it and} the ``right'' technicians
(required to achieve progress) 
as those who would be as happy in heaven as in hell, as long as they have the
tools to reach their objectives\footnote{``Mit Fernrohr und Logarithmentafel
w\"are der rechte Astronom in Himmel und H\"olle gleich zufrieden, und ebenso
der rechte Mechaniker mit Rei\ss brett und Drehbank.''}. These
words of Karl Schwarzschild, sound as true today, as they were one 
hundred years ago.
         

\end{document}